\documentclass[12pt]{book}

\usepackage[dvips]{graphicx,color}
\usepackage{makeidx,tsukuba}

\makeauthorindex
\makeindex

\begin{document}

\BookTitle{\itshape The 28th International Cosmic Ray Conference}
\CopyRight{\copyright 2003 by Universal Academy Press, Inc.}
\pagenumbering{arabic}

\chapter{Whipple Telescope Observations of Potential TeV Gamma-Ray Sources Found by the Tibet Air Shower Array}

\author{%
%
%
G.~Walker,$^{1,2}$ D.~Kieda,$^1$ I.H.~Bond, P.J.~Boyle, S.M.~Bradbury, 
J.H.~Buckley, D.~Carter-Lewis, O.~Celik, W.~Cui, M.~Daniel, M.~D'Vali,
I.de~la~Calle~Perez, C.~Duke, A.~Falcone, D.J.~Fegan, S.J.~Fegan,
J.P.~Finley, L.F.~Fortson, J.~Gaidos, S.~Gammell, K.~Gibbs,
G.H.~Gillanders, J.~Grube, J.~Hall, T.A.~Hall, D.~Hanna, A.M.~Hillas,
J.~Holder, D.~Horan, A.~Jarvis, M.~Jordan, G.E.~Kenny, M.~Kertzman,
J.~Kildea, J.~Knapp, K.~Kosack, H.~Krawczynski, F.~Krennrich,
M.J.~Lang, S.~LeBohec, E.~Linton, J.~Lloyd-Evans, A.~Milovanovic,
P.~Moriarty, D.~Muller, T.~Nagai, S.~Nolan, R.A.~Ong, R.~Pallassini,
D.~Petry, B.~Power-Mooney, J.~Quinn, M.~Quinn, K.~Ragan, P.~Rebillot,
P.T.~Reynolds, H.J.~Rose, M.~Schroedter, G.~Sembroski, S.P.~Swordy,
A.~Syson, V.V.~Vassiliev, S.P.~Wakely, T.C.~Weekes, J.~Zweerink \\
{\it
(1) Department of Physics, University of Utah, Salt Lake City, UT, USA\\
(2) The VERITAS Collaboration--see S.P.Wakely's paper} ``The VERITAS
Prototype'' {\it from these proceedings for affiliations}
}

\section*{Abstract}
An all-sky survey performed with the Tibet Air Shower Array (Tibet AS) has found a number of potential point sources of TeV gamma rays.  If they are steady sources, the implied Tibet AS fluxes should be visible with strong significance to the Whipple 10-m gamma-ray telescope (E $>$ 400 GeV) with only a short (5 hour) exposure.  We have observed four candidate directions from the Tibet-II HD dataset for $\sim$5 hours each with the Whipple telescope.  In addition, we observed a new candidate direction from the Tibet-III Phase 1 dataset for 7.5 hours.  We have found no corresponding excesses at the flux levels implied, and we have set upper limits for each candidate.

\section{Introduction}
The Tibet AS is an array of particle detectors located in Tibet at an elevation of 4300 m and operated by the TibetAS$\gamma$ collaboration.  The Tibet-II HD array operated from 1996 to 1999 with an energy threshold of 3 TeV and an angular resolution of $0.9^\circ$.  The Tibet-III array has operated since 1999 with the same threshold and resolution [1,2].  The ability of these arrays to detect gamma-ray sources has been demonstrated through detection of the Crab Nebula [3] and Mrk 501 [4].\\
\indent A wide angle survey conducted with the Tibet-II HD array found 19 directions with excesses greater than 4$\sigma$ over the average background.  The TibetAS$\gamma$ collaboration noted that these may be explained as statistical fluctuations, but one direction corresponded to the Crab Nebula [1].  We searched radio, optical, x-ray, and gamma-ray catalogues for corresponding objects within 1$^\circ$ of each of the Tibet directions, and we selected four promising candidate directions for observation with the Whipple telescope during the 2001-2002 season.  After receiving an update from the Tibet-III all-sky survey, we selected one more candidate for observation during the 2002-2003 season.  Table 1 gives a summary of the targets and observations.  The candidates were selected as follows:

\begin{itemize}
\item Tibet1 had a high significance and showed steady increase through Tibet-II HD data.
\item Tibet9 is $0.3^\circ$ from a Seyfert 1 galaxy (RGB J1337+243).
\item Tibet14 is $0.7^\circ$ from an EGRET unidentified (3EG J2021+3716).
\item Tibet16 had a high significance and is in the Cygnus star field.
\item Tibet0554 showed steady increase through Tibet-II HD and Tibet-III and was second in significance to the Crab.
\end{itemize}

\begin{table}[t]
 \caption{Tibet gamma-ray source candidates}
\begin{center}
\begin{tabular}{l|cccccc}
\hline
 &Tibet & &  & Tibet     & Whipple     & Exposure  \\
Name &Dataset & RA & Dec & Excess & Observations & (hours) \\
\hline
 *Crab&Tibet-II HD & 5h 33.2m & 22.2$^\circ$ & 4.8$\sigma$ & & \\
 Tibet1&Tibet-II HD& 3h 47.2m & 34.2$^\circ$ & 4.9$\sigma$  & Oct01 - Feb02 & 6.0  \\
Tibet9&Tibet-II HD& 13h 38.4m & 24.2$^\circ$ & 4.2$\sigma$  & Feb02 - Jun02 & 4.7  \\
 Tibet14&Tibet-II HD& 20h 21.6m & 37.9$^\circ$ & 4.2$\sigma$ & May02 - Jul02 & 4.2  \\
 Tibet16&Tibet-II HD& 21h 29.6m & 45.3$^\circ$ & 4.8$\sigma$ & Oct01 - Jun02 & 4.7  \\
\hline
*Crab&Tibet-II HD + III & 5h 34.4m & 22.0$^\circ$ & 5.4$\sigma$ & & \\
Tibet0554&Tibet-II HD + III&5h 54.8m & 30.1$^\circ$ & 4.8$\sigma$ & Dec02-Feb03 & 7.5 \\
\hline
\end{tabular}
\end{center}
\center{*Crab included for reference}
\end{table}

\section{Observations and Analysis}
Observations were made with the Whipple 10 m gamma-ray telescope with the 490 pixel camera [5].  Only the inner 379 pixels ($2.4^\circ$) were used in the analysis, and because of the large uncertainty in the Tibet AS source coordinates, 2-dimensional (2-D) analysis was required.  In the analysis, potential gamma rays were selected by applying standard supercuts shape cuts.  Then the {\it distance} and {\it alpha} cuts ($\alpha < 10^\circ$) were applied across a grid of points, with {\it alpha} and {\it distance} calculated with respect to each point.  The background was estimated using either OFF-source data, or an average background comprised of many OFF runs.  We applied this technique to data with the Crab off-axis and found it to be effective, as seen in Figure 1.

\begin{figure}[t]
  \begin{center}
    \includegraphics[height=13.5pc]{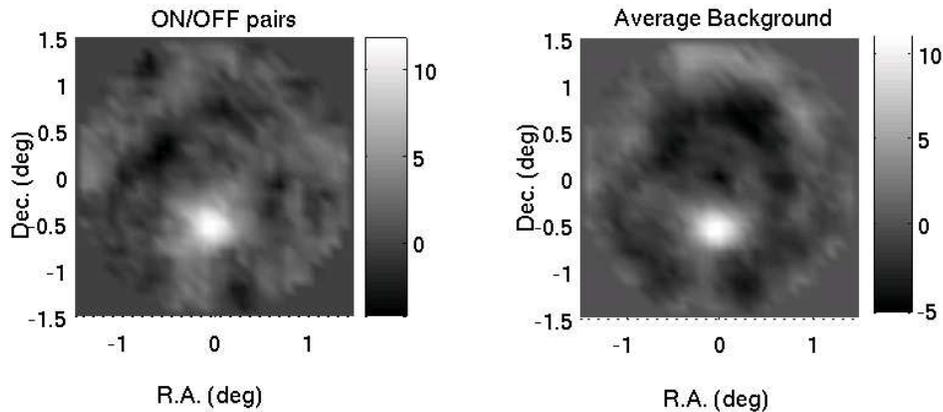}
  \end{center}
  \vspace{-0.5pc}
  \caption{Significance maps for the Crab at $0.5^\circ$ offset and 5.1 hour exposure.  Both figures were created using the same ON-source dataset, but in the left, the corresponding OFF-source data was used, while in the right an average background was used.   The scale represents pre-trials significance in $\sigma$.}
\end{figure}

\begin{table}[b]
 \caption{Results of Analysis}
\begin{center}
\begin{tabular}{l|cccc}
\hline
 & Expected & Measured & Expected Flux Ratio & Measured 2$\sigma$\\
 & Significance & Significance & To Crab Based & Upper Limit From\\
Candidate & (pre-trials) & (pre-trials) & On Tibet Data* & Whipple Data*\\
\hline
Tibet1 & 10$\sigma$ & 3$\sigma$ & 1.0 & 0.5 Crab\\
Tibet9 & 8$\sigma$ & 2.5$\sigma$ & 0.9 & 0.5 Crab\\
Tibet14 & 7$\sigma$ & 2$\sigma$ & 0.9 & 0.5 Crab\\
Tibet16 & 9$\sigma$ & 3$\sigma$ & 1.0 & 0.6 Crab\\
Tibet0554 & 9$\sigma$ & 2$\sigma$ & 0.9 & 0.4 Crab\\
\hline
\end{tabular}
\end{center}
\center{*Assuming a Crab-like spectrum}
\end{table}

\section{Results}
The 2-D analysis of the 2001-2003 Whipple data has failed to detect a source in any of the target regions.  A 2-D significance plot for Tibet0554 is shown in Figure 2 as an example.  Assuming that each candidate is point-like and has a spectral shape similar to the Crab, we are able to set upper limits on the TeV gamma-ray flux from each candidate region.  These flux limits are summarized in Table 2.

\begin{figure}[t]
  \begin{center}
    \includegraphics[height=13pc]{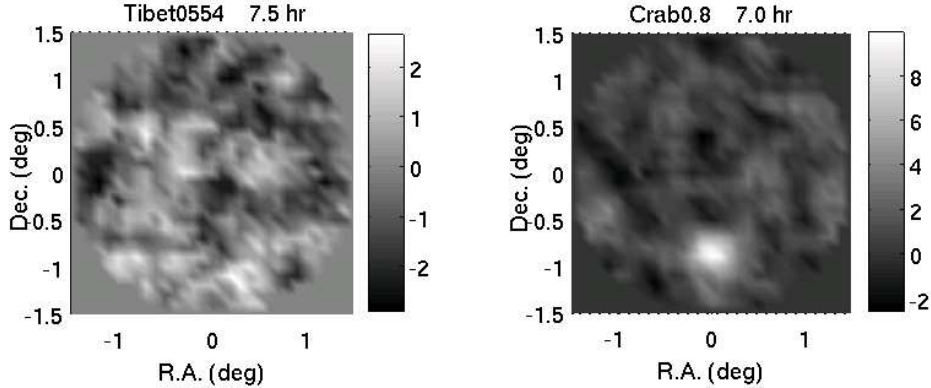}
  \end{center}
  \vspace{-0.5pc}
  \caption{Significance map for Tibet0554.  The Crab at $0.8^\circ$ offset is included for comparison.  The scale represents pre-trials significance in $\sigma$.}
\end{figure}

It should be noted that this analysis cannot rule out the possibility that the Tibet AS had observed an extended source ($size > 0.2^\circ$) or a line source with $Energy>$ 3 TeV.  However, assuming the candidates are point-like with power-law energy spectra, we conclude that the five Tibet AS excess regions we studied were either statistical fluctuations in the Tibet AS data, or due to the non-contemporaneous observation periods, the emissions were episodic.

\section{Acknowledgement}
We gratefully acknowledge support from the University of Utah and the National Science Foundation under NSF Grant \#0079704.  This research is also supported by grants from the U.S. Department of Energy, by Enterprise Ireland and by PPARC in the UK.  We also acknowledge the technical assistance of E. Roache and J. Melnick.

\section{References}

\vspace{\baselineskip}

\re
1.\ Amenomori \ et al. \ 2001, Proc. 27th Int. Cosmic Ray Conf (Hamburg), 2544
\re
2.\ Amenomori \ et al. \ 2002, ApJ, 580, 887
\re
3.\ Amenomori \ et al. \ 1999, ApJL 525, L93
\re
4.\ Amenomori \ et al. \ 1999, ApJ, 532, 302
\re
5.\ Finley, J.P.\ et al.\ 2001, Proc. 27th Int. Cosmic Ray Conf (Hamburg), 2827

\endofpaper
\end{document}